\documentclass[prl,superscriptaddress,twocolumn,showpacs,preprintnumbers,amsmath,amssymb]{revtex4}
\input psfig.sty
\usepackage{graphicx}
\usepackage{epstopdf}
\begin{document}  
\title{Scaling in soft spheres: fragility invariance on the repulsive potential softness}
\author{Cristiano De Michele${}^{1,2}$}
\noaffiliation
\author{Francesco Sciortino}
\affiliation{Dipartimento di Fisica, INFM and 
INFM Center "SOFT: Complex Dynamics in Structured Systems"   ,
Universit\'a di Roma "La Sapienza", 
Piazzale Aldo Moro 2, I-00185  Roma, Italy}
\author{Antonio Coniglio}
\affiliation{Dipartimento di Scienze Fisiche and INFM-Coherentia,
Universi\'a di Napoli "Federico II", 
Via Cinthia (Monte S. Angelo) Ed. G, I-80126 Napoli, Italy}
\date{\today}
\begin{abstract}
 We address the question of the dependence of the fragility of glass forming supercooled  liquids  on the ''softness'' of an interacting potential  by performing 
numerical simulation of a binary mixture of soft spheres with different power $n$ of the 
interparticle repulsive potential. We show that the temperature dependence of the diffusion coefficients for various $n$ collapses onto a universal curve, supporting the unexpected 
view that fragility is not  related to the hard core repulsion. We  also find that the
configurational entropy correlates with the slowing down of the dynamics for all studied $n$. 
\end{abstract}
\maketitle
When a liquid is cooled below its melting temperature, if crystallization does not take place, it becomes {\it supercooled}. In this supercooled region, the viscosity increases by more than 15 order of magnitude in a  a small $T$-range. When the viscosity $\eta$ reaches a value of about $10^{13}$ Poise  the liquid can be treated as an amorphous solid, i.e., a glass \cite{stillnaturerev,torquatorevglass,stillglassreviewscience,angellnatureliqland}
and the corresponding temperature is defined as  the glass transition temperature (labelled $T_g$).

The $T$-dependence of the viscosity $\eta$ differs for different glass
formers. Angell has proposed a classification based on the behaviour of $\eta(T)$.
Glasses are said to be {\it fragile} if they show large deviations  from an Arrhenius law ($\eta(T)\propto \exp[E/T]$) or {\it strong} otherwise  \cite{angellfrag}. The fragility $m$ of a glass forming liquid can be quantified by the slope of  $\log\eta(T)$ vs $T_g/T$, evaluated at $T_g$, i.e. as
\begin{equation}
m =       \frac{{\it d}\log \eta}{{\it d}(T_g/T)}|_{T=Tg}
\label{eq:fragdlog}
\end{equation}
While the original definition of fragility is based on a purely  dynamic quantity, correlation between $m$ and other physical properties of glass forming liquids, both with dynamic and thermodynamic properties,  have been reported . Recently, a remarkable correlation with vibrational properties of the glass state has been discovered \cite{tullioscience}.
One of the main challenges in the physics of supercooled liquid and glasses 
is to understand the connection between dynamical properties of the liquid close to the glass transition, i.e. the fragility,  and microscopic properties. Is the fragility most affected by the steepness  of the repulsive potential or by the  inter particle attraction? Is it controlled by other properties of the interaction potential? In the present letter we address this question
calculating numerically the fragility of several models for liquids, differing only 
in the softness of the repulsive potential.  We aim at understanding whether changing the softness of the repulsive potential  the fragility changes accordingly.
We show more generally that the diffusion coefficient $D$ can be scaled on a universal master
curve by changing the softness of the repulsive potential. This implies that the fragility
does not depend on the softness of the interaction potential.
We complement this dynamical study with the evaluation of the configurational
entropy to check the validity of the Adam-Gibbs\cite{AG,antosconf,schillingreview,wolynes} relation.

In this Letter we consider a simple glass former, a $80:20$ binary mixture of $N=1000$ soft spheres \cite{replicaBMLJPRL,speedySSPEL,depablosoft},
that is an ensemble of spheres interacting via the following potential
\begin{equation}
V_{\alpha\beta}(r) = 4\epsilon_{\alpha\beta} \left(\frac{\sigma_{\alpha\beta}}{r}\right)^n
\label{Eq:Vsoftsphere}
\end{equation}
where $\alpha,\beta\in{A,B}$, $\sigma_{AA}=1.0$, $\sigma_{AB} = 0.8$, $\sigma_{BB} = 0.88$,
$\epsilon_{AA}=1.0$, $\epsilon_{AB}=1.5$, $\epsilon_{BB}=0.5$ and
$n$ is a parameter by which is possible to tune the ''softness'' of the interaction
\cite{hansenmcdonald}. This interaction potential is a Kob-Andersen potential 
\cite{kobandersenPRE} in which  the attractive part of the potential has been dropped.
In particular we investigate the values $n=6,8,12,18$.

%
This choice for the binary mixture is motivated by the fact that such a system is  
not prone to crystallization,  that is it can be easily supercooled below its melting temperature
. Still, for $n < 6$, crystallization takes place within the simulation time, determining a lower limit
to the range of investigated $n$ values. 
Reduced units will be used in the following, length will be in units of $\sigma_{AA}$,
energy in units of $\epsilon_{AA}$ and time in units of 
$(M\sigma_{AA}^2/\epsilon_{AA})^{1/2}$, where $M$ is the mass of all particles. 
In physical units, assuming the atom $A$ is Argon, the units are a length of  $3.4$$\AA$, an energy of $120K k_B$ and a time of $2.15 ps$.

The self-similar nature of the soft-sphere potential couples $T$ and $V$. It can be shown that all thermodynamic properties depend on the quantity $TV^{\frac{1}{n}}$\cite{softeos}. Dynamic properties can also be scaled accordingly \cite{japsoft}. Hence, it is sufficient to quantify the 
$T$-dependence or the $V$-dependence of any observable to fully characterize the behavior of the system.  As a consequence the fragility does not change
upon changing the density of the soft binary mixture.

\begin{figure}
\begin{center}
\psfig{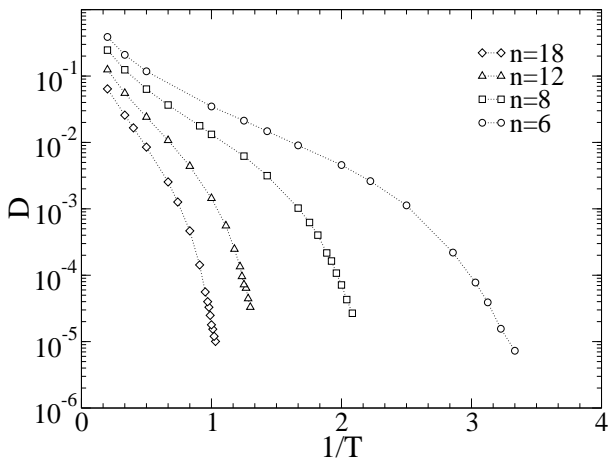}
\label{Fig:DallMCT}
\end{center}
\caption{
$T$-dependence of the 
diffusion coefficient for all $n$ investigated.
$D$ is measured in units $\sqrt{\sigma_{AA}^2 \epsilon_{AA}/ M}$. }
\label{Fig:diffMCT}
\end{figure}

Figure \ref{Fig:diffMCT} shows the $T$-dependence of the diffusion coefficients, evaluated from the long time limit behavior of the mean square displacement,  for all $n$
investigated and covering a  window of about four order of magnitudes.

\begin{figure}
\begin{center}
\psfig{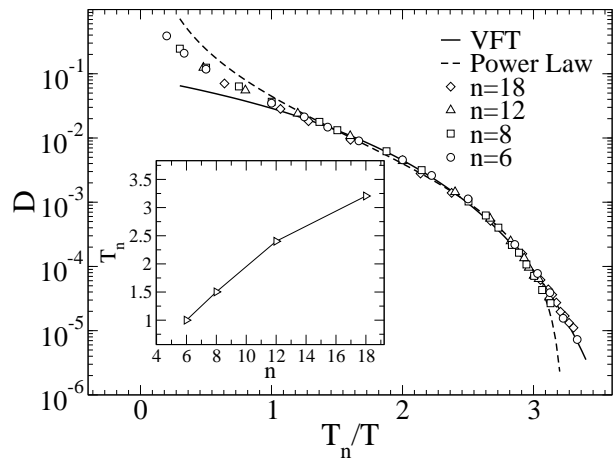}
\end{center}
\caption{ 
Master curve for diffusion coefficient $D$ calculated rescaling the temperature 
by an $n$-dependent quantity.
Full line is a Vogel-Fulcher-Tamman fit to all points lying in the landscape-influenced region 
\cite{sastryPELdominated,sastryPELdynregimes}, which for the present model corresponds to
$\xi < 0.78 $ with $\xi = T/T_n$,  
, i.e. ${\cal D}(\xi)=A\exp[B/(\xi-\xi_0)]$.  The fitting parameters $\ln A$, $B$ and $\xi_0$ are respectively $-2.46$,$-0.857$,$0.209$.  From this master curve one can also estimate the dynamic fragility, that is about $130.4$. Dashed line is a fit to all points lying in the interval 
$ 0.328 < \xi <  0.833 $,  to a power law \cite{goetzepisa,goetze,sastrylowTBMLJ}, i.e. ${\cal D}(\xi)=A_m (\xi-\xi_c)^\gamma$, where $A_m=-2.41$,$\xi_c=0.309$,$\gamma=1.88$. The inset shows the 
$n$-dependence of the scaling parameter $T_n$.}
\label{Fig:scaleDT}
\end{figure}

In the attempt of compare the $n$-dependence of the diffusion
coefficient, we report in Fig.  \ref{Fig:scaleDT} the data as a function of $T_n/T$, where $T_n$
is chosen in such a way to maximize the overlap between data of
different $n$, i.e. to collapse all data onto a single master
curve. Figure \ref{Fig:scaleDT} shows that all curves can be
successfully scaled onto the master curve ${\cal D}$ choosing a proper
set of scaling parameters $T_n$ (whose $n$-dependence is plotted in
the inset of this Figure). The very good quality of the resulting
master curve 
\begin{equation}
D(T) = {\cal D}(T/T_n).
\label{Eq:Dscaling}
\end{equation}
suggests that the $n$-dependence enters only via a rescaling of the temperature \cite{note3,tarjusmossa}.
The remarkable consequence of latter result is that the fragility of the system does not 
depend on the repulsive interaction potential. In fact according to Eq.~(\ref{Eq:Dscaling}) and
from the definition of liquid's fragility $m$ given in Eq.~(\ref{eq:fragdlog}),
assuming $D\propto\tau^{-1}$ we get:
\begin{equation}
m = \frac{T_g(n)}{T_n}\; \frac{1}{{\cal D}[T_g(n)/T_n]}\;
\frac{d{\cal D}(x)}{dx}|_{x=T_g(n)/T_n}
\label{Eq:scaledfrag}
\end{equation}
where $T_g(n)$ is the glass transition temperature for the system with softness
$n$, which can be defined as the temperature at which diffusivity reaches  an arbitrary small value  $10^{\cal K}$\cite{note}, i.e.
\begin{equation}
-\log D[T_g(n)] = -\log {\cal D}\left [ \frac{T_g(n)}{T_n}\right ]={\cal K}
\label{Eq:BSdefTg}
\end{equation}

Eq.\ref{Eq:scaledfrag}  shows that the fragility index $m$ is a function only of the scaled variable 
$\frac{T_g(n)}{T_n}$ and hence, as far as the scaling reported in Fig. \ref{Fig:scaleDT} keeps holding even at temperatures lower than the one we are able to equilibrate, the dynamic fragility $m$ is independent of $n$ as well. By fitting the master curve to a  Vogel-Tamman-Fulcher fit, as shown in Fig. \ref{Fig:scaleDT}, an estimate of $\frac{T_g(n)}{T_n} = 10^{\cal K} $ can be calculated, resulting into a
estimation of $m \approx 130$. This figure should be compared with the value $m=81$ for 
o-terphenyl (OTP), that is a typical fragile liquid and $m=20$ for the prototypical strong glass the 
liquid silica ($SiO_2$). 

For completeness, we report also in Fig. \ref{Fig:scaleDT} a fit of the master curve according to the prediction of  mode-coupling theory, which has been shown to be consistent with numerical data for several models in the weak supercooling region.   A best fit procedure requires the exclusion of the low $T$ points, for which deviations from the power-law fit are observed 
\cite{note2}.

\begin{figure}
\begin{center}
\psfig{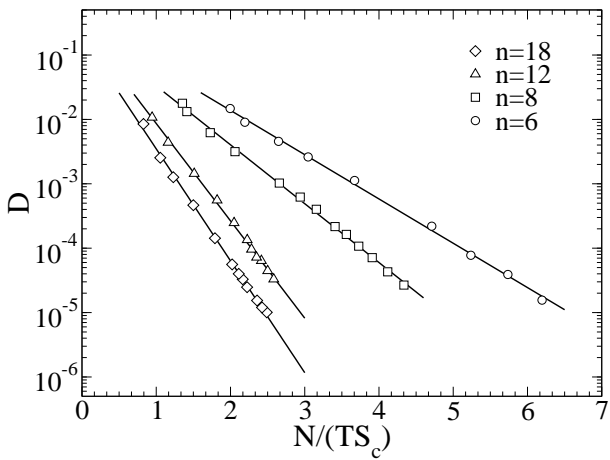}
\label{Fig:AGDTanh}
\end{center}
\caption{%
Test of the Adam-Gibbs relation based on the configurational
entropy values calculated subtracting the anharmonic and harmonic entropies to 
the total entropy (see \cite{crifraJCP} for details). The fitting parameters
$\ln A_{AG}$ and $B_{AG}$ are: 
for $n=6$, $-1.112$, $-1.584$;
for $n=8$, $-1.303$,  $-2.105$ ;
for $n=12$, $-1.271$, $-3.481$;
for $n=18$, $-1.657$, $-4.002$.
$TS_c$ is measured in units of $\epsilon_{AA}$.
}
\label{Fig:AGallANH}
\end{figure}

Recently, evidence has been presented that kinetic fragility strongly correlates with
thermodynamic fragility\cite{angelmartinez}. In this respect, it is worth looking if
the scaling observed in dynamical properties has a counterpart in thermodynamic properties.
In particular, we  evaluate the configurational entropy for 
the system,  within the potential energy landscape framework as discussed in details in Refs.\cite{crifraJCP,fraPELPRL,stillweberPRA,emiliaOTP,sastryPELform,sastryBMLJnature}.  In brief, we estimate $S_{c}$ as difference between the
liquid entropy (calculated via thermodynamic integration from the ideal gas) and of the
vibrational entropy (calculated via thermodynamic integration, including anharmonic corrections, from the very low temperature harmonic dynamics of the disordered solid associated to the liquid configuration).  We then focus on the ability of the Adam-Gibbs (AG) relation --- which states that
\begin{equation}
D(T) = A_{AG} e^{\frac{B_{AG}}{T S_c}},
\label{Eq:AdamGibbs}
\end{equation}
--- of modelling the temperature dependence of $D$.   Fig.~\ref{Fig:AGallANH}  shows the AG plot for the studied $n$ values.  For all $n$, a satisfactory linear representation of $\log(D)$ vs. $1/TS_{c}(T)$ is observed.  As discussed in more details in Ref.\cite{jchemphysme}, the simultaneous validity of the VTF description of $D$ and of the AG relation requires the identity of the kinetic and thermodynamic fragilities. In this respect, the independence of $n$ discussed above for the case of kinetic fragility carries on also to thermodynamic fragility.

\begin{figure}
\begin{center}
\psfig{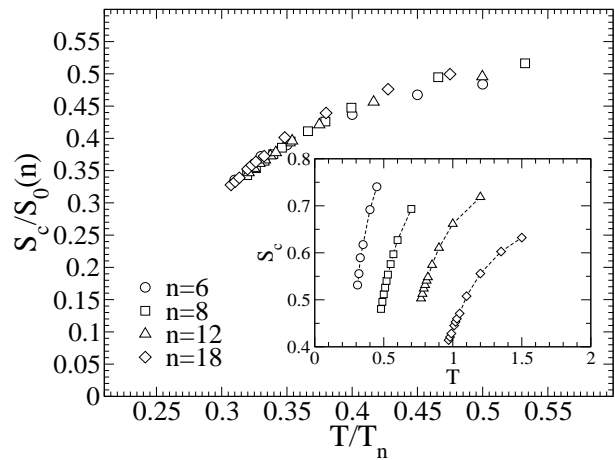}
\end{center}
\caption{%
Master plot of the configurational entropy  (see Eq.~(\ref{Eq:Scscaled})). 
The inset shows the temperature dependence of the configurational entropy.}%
\label{Fig:Scscaled}
\end{figure}

A remarkable consequence of the validity of the AG relation (Eq. \ref{Eq:AdamGibbs}), associated to the scaling with $n$ of $D$ (Eq.\ref{Eq:Dscaling}) is that 
the configurational entropy can be written as 
\begin{equation}
S_c(T) = S_0(n) {\cal F}(T/T_n) 
\label{Eq:Scscaled}
\end{equation}
where $F(x)$ is a scaling function and $S_0(n) = B_{AG}/T_n$. To support such
proposition, we show in  Figure \ref{Fig:Scscaled}   $S_c$ multiplied by the factor $B_{AG}/T_n$ as a function of  $T/T_n$, were $T_n$ are the values for which $D$ scaling is recovered 
(inset of Fig.\ref{Fig:scaleDT}). Again, the quality of the data collapse stresses the validity of the
scaling with $n$. 

To conclude the relevant result that has been shown in this Letter is that in the case of 
soft sphere potentials,  the dynamic fragility  is independent on the power $n$ of the
short range repulsion.    This conclusion is based on the hypothesis that the scaling observed in the range of $T$ where simulations are feasible extends also to lower temperatures, down to the glass transition temperature. Indeed, a particular effort has been made to equilibrate configurations to temperatures lower than the MCT temperature, where dynamical processes different from the ones captured by MCT are active.  If the scaling is indeed valid, 
the results presented in this Letter strongly support the possibility, that contrary to our common understanding, fragility in liquids is mostly controlled by other properties of the potential, more than by the  hard core repulsion.  
Finally we note that one could be tempted to associate the fact that the diffusion coefficient
data can be rescaled only by change the energy scale by $T_n$ to a simple overall rescaling of the landscape potential. The data in Fig.\ref{Fig:Scscaled}  suggest that this is not the case since $S_c(E)$
is not just scaling function of $T/T_n$ but it needs to be rescaled by a factor $S_0(n)$ 
and hence the number of distinct basins explored 
at the same $T/T_n$ changes with $n$.
A non-trivial compensation mechanism between the scaling of the static properties ($S_c$) and
the scaling of the kinetic coefficient $B_{AG}(n)$ (defined in Eq. (\ref{Eq:AdamGibbs})) on $n$ must be present.

We thank L.~Angelani and G.~Ruocco for useful discussions. We acknowledge support from INFM Initiative Parallel Computing, Marie Curie Network and Miur FIRB and COFIN2002.


\begin{thebibliography}{40}
\bibitem{note3} Note that linear extrapolation of $T_n$ with $n$ shows that $T_n$ goes 
to $0$ at same value $n\simeq 2$. This may suggest that below a critical value of the range of the potential the $T$-dependence of diffusivity coefficients exhibits a strong crossover to a different regime. Below this critical value the $T$-dependence of the diffusion coefficient
should be weak.
\bibitem{note} In particular we have made the choice $D[T_g(n)]=10^{\cal K} = 5.75\times 10^{-16}$, 
this value for the diffusion coefficient ensures that at $T_g$ 
the relaxation time is about $100 s$.
\bibitem{note2} We also note that a Bassler form ($D(\xi) = A \exp (B/\xi^2)$) does not reproduce the data in a manner comparable to VFT and MCT.
\end{thebibliography}
\end{document}